\documentclass[prl,twocolumn]{revtex4}

\usepackage{epsfig}
\usepackage{latexsym}
\usepackage{bm}

\newcommand{\be}{\begin{equation}}
\newcommand{\ee}{\end{equation}}
\newcommand{\bea}{\begin{eqnarray}}
\newcommand{\eea}{\end{eqnarray}}

\def\beginwide{
        \end{multicols} \vspace*{-0.5cm} \noindent
        \rule{3.5in}{.1mm}\rule{.1mm}{5mm} \widetext \medskip }
\def\beginwidetop{
        \end{multicols} \vspace*{-0.5cm} \noindent
        \widetext \medskip }
\def\endwide{
        \hspace*{3.35in}~\rule[-5mm]{.1mm}{5mm}\rule{3.5in}{.1mm}
        \begin{multicols}{2} \vspace*{-1.0cm} \noindent }
\def\endwidebottom{
        \begin{multicols}{2} \vspace*{-1.0cm} \noindent }

\begin{document}

\title{Sampling rare  fluctuations of height in the Oslo ricepile model}

\author{Punyabrata Pradhan and Deepak Dhar} 
\address{Department of Theoretical Physics, Tata Institute of Fundamental
Research, Homi Bhabha Road, Mumbai-400005, India}

\begin{abstract} 

We have studied large deviations of the height of the pile from its mean
value in the Oslo ricepile model. We sampled these very rare events with
probabilities of order $10^{-100}$ by Monte Carlo simulations using
importance sampling. These simulations check our qualitative arguement
[Phys. Rev. E, {\bf 73}, 021303, 2006] that in steady state of the Oslo
ricepile model, the probability of large negative height fluctuations
$\Delta h=-\alpha L$ about the mean varies as $\exp(-\kappa {\alpha}^4
L^3)$ as $L \rightarrow \infty$ with $\alpha $ held fixed, and   
$\kappa > 0$.

\vspace{0.1cm}
\typeout{polish abstract} 

\end{abstract}

\maketitle

\section{1. Introduction}

Large deviations of fluctuations in a system have been studied extensively
\cite{Lanford, Evans} and recently have attracted much attention
especially in the non-equilibrium stationary states of driven systems
\cite{Bertini, derrida, stinchcombe, kurchan}. In a recent paper, we have
argued that in the critical slope type stochastic toppling models in $d$
dimensions, the probability of large negative height fluctuations $\Delta
h=-\alpha L$ about the mean for a system of size $L$ decays
superexponentially, as $\exp( - \kappa {\alpha}^{\frac{d+2}{1-\omega}}
L^{d+2})$ , for $L \rightarrow \infty$, with $\alpha > 0$ held fixed
\cite{pradhan-dhar}. Here $\kappa > 0$   and $\omega$ is
an exponent defined by $\langle (\Delta h) ^2 \rangle \sim L^{2\omega}$.
Since the arguements are plausible but not rigorous, it seems worthwhile
to check them by numerical simulations. However, straight-forward sampling
techniques fail in this case, as the probabilities become very small even
for fairly small $L$. For example, for $L= 10$, in $d =1$, already the
probability of the minimum slope configuration is of order $10^{-45}$.

In this paper, we numerically estimate the probabilities of large negative
height fluctuations of the pile in the steady state in the one dimensional
Oslo ricepile model using  a variation of ``go with the winners"
strategy  \cite{grassberger}. We estimate the full probability
distribution function $Prob_L(\Delta h)$ where $\Delta h$ is height
fluctuations of the pile about its mean value.  This distribution has a
scaling form $L^{-\omega} g(\Delta h/L^{\omega})$. The $1d$ Oslo model is
expected to be the Edwards-Wilkinson universality class, for which $\omega
= 1/4$ \cite{pruessner}.  Our non-rigorous arguments in \cite{pradhan-dhar} 
imply that the scaling function  should vary  as $\exp(-\kappa x^{4})$ for 
$x \ll -1$ where  $\kappa > 0$. Our numerical data fully supports the 
theoretical expectation.

\section{2. Definition of the Model}

The Oslo ricepile model \cite{frette, oslomodel} is a stochastic
sandpile-like model defined on a one dimensional lattice with a critical
threshold value for the slope above which a toppling occurs, and the
threshold is randomly reset after each toppling.  Here we use an
equivalent version of the rules as given in \cite{dhar1}: We consider a
chain of lengh $L$. A configuration of the pile is specified by an integer
height variable $h_i$ at each site $i$. The slope $z_i$ at site $i$ is
defined to be $h_i-h_{i+1}$, with $h_{L+1} \equiv 0$ . Any site $i$ with
slope $z_i=1$ is stable. Any site $i$ with slope $z_i \ge 3$ is said to be
unstable, and relaxes by toppling.  Slope $2$ can be either stable or
unstable. Whenever slope at a site reaches the value $2$ from a different
value, because of incoming or outgoing grains, it is created as an
unstable $2$ (denoted by $\bar{2}$).  A $\bar{2}$ becomes stable $2$  
(denoted by $2$ without overbar) with probability $p$ without any toppling,
or it topples with probability $q = 1-p$.  Whenever there is a toppling at
site $i$, one grain is moved from the site $i$ to $i+1$.  If there is a
toppling at the right end $i=L$, one grain goes out of the system.  
Grains are added only at site $1$ and  only after avalanche caused by the 
previous grain has stopped.

The $1d$ Oslo ricepile model has a remarkable Abelian property that the
final height configuration does not depend on the order we topple the
unstable sites \cite{dhar1}. Also, after addition of total $L(L+1)$ grains
to any configuration, probabilities of different stable configurations are
exactly the same as in the steady state, independent of the initial
configuration \cite{dhar1}. The number of recurrent stable configurations
in the critical states can be calculated exactly, and is approximately
$\frac{1}{\sqrt{5}}(\frac{1+\sqrt{5}}{2})^{2L+1}$ for large $L$
\cite{chua}. In the steady state height profile always fluctuate between
slope $1$ and $2$. The height $h_1$ at the site $1$ has a stationary
probability distribution, $Prob_L(h_1)$, which is sharply peaked near its
average value $\bar{h}_1$.  For large system size $L$, the average height
$\bar{h}_1$ varies linearly with $L$, and and the fluctuations in $h_1$
scale as a sublinear power of $L$, with variance of $h_1$ varying as $L^{2
\omega}$, with $0 < \omega < 1$.

\section{3. Exact calculation of $Prob_L(h_1)$  for small $L$.}

The probability distribution of $h_1$ in the steady state can be exactly
calculated numerically for small $L$ using the operator algebra satisfied
by addition operators \cite{dhar1}.  We recapitulate this briefly here. We
denote any configuration by specifying slope values at all sites from
$i=1$ to $i=L$ by a string of $L$ integers (with or without overbar), {\it
e.g.}, $|10\dots\bar{2}2 \rangle$.  For unstable site $1<i<L$, the rules are
as given below.

$$
|\dots z_{i-1}, \bar{2}, z_{i+1}\dots \rangle 
$$
$$
\rightarrow 
p|\dots z_{i-1}, 2, z_{i+1} \dots \rangle + 
q |\dots \overline{(z_{i-1}+1)}, 0, \overline{(z_{i+1}+1)} \dots \rangle
$$
\begin{equation}
|\dots z_{i-1}, {\bar 3}, z_{i+1}\dots \rangle 
\rightarrow 
|\dots \overline{(z_{i-1}+1)}, 1, \overline{(z_{i+1}+1)} \dots \rangle
\end{equation}
with the convention that ${\bar 1} =1$. 
At the left end, rules are as given above except that there is no left 
neighbour of site $1$. At the right end $i=L$, the rules are
$$
|\dots z_{L-1}, \bar{2} \rangle \rightarrow p|\dots z_{L-1}, 2\rangle + 
q|\dots \overline{(z_{L-1}+1)}, 1\rangle 
$$
\begin{equation}
|\dots z_{L-1}, 3 \rangle \rightarrow |\dots \overline{ z_{L-1}+1}, 
\bar{2}\rangle 
\end{equation}

Using these  toppling rules repeatedly and the Abelian property,  we can 
relax any unstable configuration. 

Let us now consider the state $|\bar{2}\bar{2} \dots \bar{2}\bar{2} \rangle $
where all the  slopes  are unstable $2$'s. If we add one more 
grain at site $i=1$ in this state, we get the same state back (toppling the
site with $z_i=3$ repeatedly) which implies that it is the steady state. So if 
we relax this configuration fully, we get probabilities of all the 
configuration in the steady state. For example, if we relax 
$|\bar{2}\bar{2}\rangle$ for $L=2$, we get the following sequence, 
\[|\bar{2}\bar{2}\rangle 
\rightarrow
p|2\bar{2}\rangle + q|1\bar{2}\rangle
\rightarrow
\]
\[
p^2|22\rangle+pq|12\rangle+pq|1\bar{2}\rangle+q^2|\bar{2}1\rangle
\rightarrow 
\dots \rightarrow
\]
\[
p^2|22\rangle+(p+p^2)q|12\rangle+(p+p^2)q^2|21\rangle
\]
\vspace{-0.2cm}
\[
\mbox{~~~~~~~~~~~~~~~~~~~~~~~~~~~~~~~~}+
(p+p^2)q^3|02\rangle+(1+p)q^4|11\rangle .
\]

\begin{figure}
\begin{center}
\leavevmode
\includegraphics[width=9.0cm,angle=0]{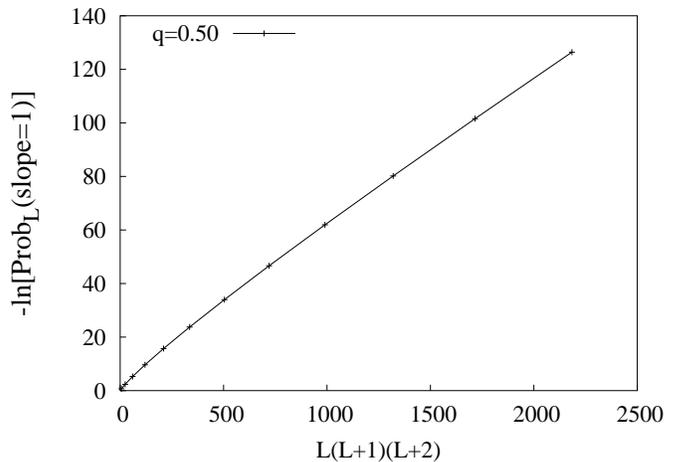}
\caption{\small (Color online) Logarithm (base $10$) of  the probability 
of occurrence of the minimum slope configuration, calculated  exactly 
numerically, is plotted against system size $L(L+1)(L+2)$ in  the Oslo 
ricepile model.}
%\label{CONFIGURATION}
\end{center}
\end{figure}

One can similarly calculate the steady state probabilities for higher $L$.
In Table \ref{table}, we list the resulting expression for the probability
of the minimum slope configuration $|11\dots11\rangle$ for  $L=2,3,4$. 
For larger $L$, the calculation becomes very tedious.  There are two
branches for relaxing any unstable site with $z=\bar{2}$ and the total CPU
time increases as $\exp(L^3)$ as ${\cal{O}}(L^3)$ relaxations of unstable
sites are required to reach the steady state \cite{footnote1}.  We have 
calculated the probability of minimum
slope exactly numerically for $L \le 12$ using a simple code written in
$C$. We used specific numerical values $p= q = 1/2$, to simplify the
calculation, so that the probabilities are simple numbers and not
polynomials in $p$.  Even then, for $L > 12$, the computer time
required becomes prohibitive.

\begin{table}
\begin{center}
\begin{tabular}{|c|c | }

\hline
 L & Probability of the minimum slope \\
\hline
$2$ & $(1+p)q^4$  \\
\hline
$3$ & $(1+4p+6p^2+5p^3+2p^4)q^{10}$ \\
\hline
$4$ & $(1 + 10p + 45p^2 + 125p^3 + 241p^4 + 81p^9 + 18p^{10}) q^{20}$ \\
\hline

\end{tabular}
\end{center}
\caption{Table for the probability of the minimum slope for $L=2, 3, 4$.}
\label{table}
\end{table}

For large negative deviations, the probability becomes very small. Even
for system size as small as $L=6$ and $L=7$, the probabilities of the
minimum slope are $4.81 \times 10^{-11}$ and $1.76
\times 10^{-15}$ respectively. For $L=12$, this probability is $1.23
\times 10^{-55}$. We argued in \cite{pradhan-dhar} that this probability 
is ${\cal O}(q^{m})$ with $m = L(L+1)(L+2)/6$, for $q \rightarrow 0$, and is
expected to decrease as $\exp ( -\kappa(q) L^3)$ for all $q$.  In Fig.1 we
have plotted negative of logarithm of the probability of the minimum slope
configuration versus $L(L+1)(L+2)$ for $q = 1/2$. The linear increase is
in agreement with the theoretical expectation.

\section{4. Monte Carlo simulation with biased sampling.}

In the simplest Monte Carlo algorithm to estimate probabilities of
low-slope configurations in the steady state, one can start with the
unstable configuration $|\bar{2}\bar{2} \dots \bar{2}\bar{2} \rangle$, and
relax it by using the probabilistic toppling rules until a final stable
configuration is reached. If this is repeated many times, the fractional
number of configurations with a given value of $h_1$ gives an estimate of
the corresponding probability. However, this method is clearly unsuitable
for estimating probabilities which are much smaller than $10^{-10}$. For
estimating quantities like the probability of the minimal slope
configuration in the steady state, this method is useless for $L > 6$ or
so.

 Clearly, we need to implement some sort of importance sampling. In the
simplest implementation, one thinks of different possible configurations
in the course of evolution of an avalanche at each step $t$ as a branching
tree. If we reach a configuration ${\cal C}_t$ at a node at the $t$-th
step, the probability of the process stopping is, say $a({\cal C}_t)$.  
If we want to sample the low slope configurations, we do not want the
process to die too soon. Then we do not select any of the nodes that
correspond to stable configurations, and select one of the remaining
branches with probability equal to their original probability, divided by
the factor $[1-a({\cal C}_t)]$, and the final survival probability is
estimated by product of such factors.

However, this procedure also is not satisfactory for our problem as there
are some unstable nodes all of whose possible resulting stable 
configurations
have heights greater than the minimum height. For example let us consider
a case for $L=5$ where $|2110\bar{2}\rangle$ has two descendants
$|21102\rangle$ and $|21111\rangle$, both stable. In the branching tree, these 
are like dead-end streets. The relaxation process will eventually die if we
encounter with such unstable configurations. But it is difficult to
identify these directly and avoid them, without a computationally
expensive depth-search. So the resulting process still has a nonzero
probability of reaching such a node at the next step, and the overall
probability of survival still decreases exponentially with the depth of
the tree.

We now describe an algorithm that does manage to avoid this problem, but
at the cost of having to define and update one additional  real
variable at each site of the lattice.

\subsection{Algorithm for sampling rare events}

We start with a configuration with all sites unstable, and all $z_i =
{\bar 2}$. We imagine that there is a random number, uniformly distributed
between $0$ and $1$, at each site $i$.  After each toppling, the random
number at the site is replaced by a new random number.  Let the value of
this random number at site $i$ at the end of the update-step $t$ be
$x(i,t)$. At $t=0$, all $x(i,0) $ are independent random variables lying
between $0$ and $1$.

As it does not matter in which order we topple the unstable sites, we
choose the following rule: at any time step, we first topple any unstable
sites with $z_i = 3$, and reset the random number at that site.  When all
the unstable sites are with slope $2$, {\it we topple the site having the
largest random number}, if the random number is greater than $p$.  If the
number is less than $p$, the avalanche stops.  This constitutes the end of
one update step. Note that one update step may involve more than one
topplings at sites with $z =3$, but there is exactly one toppling at a
site with $z =\bar{2}$ in each update step. 

The problem is to determine the conditional probability of further
evolutions for different avalanche histories, if we impose the condition
that an avalanche does not stop.  Note that our rule of selecting the
unstable $\bar{2} $ for toppling introduces correlations between different
$x(i,t)$:  if we know the random number at the site selected for toppling,
random number at all other not-selected sites must be smaller.  Our
algorithm uses this correlation. We do not start by assigning specific
values to the random number $x(i,t)$ at all sites. The only information
about $x(i,t)$ which is known, and regularly updated, is the largest
allowed value of $x(i,t)$ determined by the known history of the
avalanche. Let us call this $x_{max}(i,t)$.

The toppling history can be specified by giving the site with $z=\bar{2}$
selected for toppling at each update step, and the random number at that
site at the time of toppling.  Let $j(t)$ be the site selected for
toppling at the $t$-th update step, and $y(t)$ be the value of the random
number at $j(t)$ at the time, i.e. $ y(t)= x(j(t),t)$.  We shall denote
this sequence upto time $t$ by ${\cal T}_t = \{ [j(r), y(r)] : r = 1 {\rm
~to~} t\}$.  Given the sites that have been toppled, one can determine the
set of unstable sites ${\bf u}(t)$ with $z = \bar{2}$, out of which the
site with the maximum random number has to be selected at update-step
$t+1$.  Since we know that $x(j,t)$ for any site $j \in {\bf u}$ has not
been selected for toppling earlier since it was reset, it must be smaller
than all the corresponding $y$'s selected since then. If it was reset at
update step $t_{prev}(j,t)$, we must have 
\begin{equation} 
x_{max}(j,t) =
Min \{y(t^{\prime}): t_{prev}(j,t) <  t^{\prime} \leq t\}.  
\end{equation}

At any update step $t+1$, we first topple any sites with $z=3$, and reset
$x_{max}$ at the site to $1$, and continue this till no slopes are $3$.  
Let ${\bf u}(t)$ be the set of unstable sites with $z=\bar{2}$. It is
straight forward to determine the conditional joint probability
distribution of $[j(t+1), y(t+1)]$, given the condition that the avalanche
does not stop, using the information in ${\cal T}_{t}$ (see below).  We
then select one of these unstable sites in ${\bf u}(t)$ as the one having
the largest random number, and assign the value of the random number at
this site using the correct conditional probability distribution.

The conditional probability distribution that $x(j,t)$ is $\leq x $ ,
given the value of $x_{max}(j,t)$, is \begin{equation} {\rm Prob}(x(j,t)
\leq x| x_{max}(j,t)) = g(\frac{x}{x_{max}(j,t)})  \end{equation} where
$g(\xi)$ as $g(\xi) = \xi$, for $ 0 \leq \xi \leq 1$, and $g(\xi)= 1$ for
$\xi >1$. As there is no correlation between the values $x(j,t)$ for
different $j$'s for the same time $t$, beyond that implied by the
conditions that $x(j,t) \leq x_{max}(j,t)$, we must have \begin{equation}
{\rm Prob}(y(t+1) \leq y|{\cal T}_t) = \Phi_t(y)= \prod_{j \in {\bf u}(t)}
g(\frac{y}{x_{max}(j,t)})  \end{equation}

If we put an additional condition that $y(t+1)>p$, the corresponding
conditional distribution of $y(t+1)$ is given by \begin{equation} {\rm
Prob}(y(t+1) \leq y|{\cal T}_t, y(t+1) \ge p) = {\frac {\Phi_t(y)  -
\Phi_t(p)}{1 -\Phi_t(p)}} \label{eq6} \end{equation}
 In the appendix we describe the algorithm for generating a random number 
with  a given distribution of the form Eq.(\ref{eq6}).

Let $F(t+1)$ be the probability that $y(t+1) \ge p$. Clearly, we have
$F(t+1)$ equals to $[1-\Phi_t(p)]$. i.e., \begin{equation} F(t+1) = 1 -
\prod_{j \in {\bf u}(t)} g(\frac{p}{x_{max}(j,t)}) \label{eq7}
\end{equation} The relative weight of a particular history ${\cal T}_t$
being realized without the avalanche getting stopped is $\prod_{t'=1}^{t}
F(t')$. We calculate the attrition factor $F(t+1)$ using Eq.(\ref{eq7}).
We then topple at the selected site $j(t+1)$, and update the values of
$x_{max}(j(t+1))= 1$, and set $x_{max}(j')= y(t+1)$ for $j' \neq j$. And
repeat.

After we start relaxing the unstable configuration $|\bar{2}\bar{2}
\bar{2} \dots \dots \bar{2} \rangle$, the height at site $1$ gradually
decreases. At some step of relaxation, the height at first site becomes
$h_1 \le h$ for the first time in the course of relaxation. We multiply
all the previous factors, $F(t)$'s, upto this step and this product gives
\begin{equation} W(h) = \prod_{\{{\cal C}_t: h_1 > h\}} F(t)
\end{equation} Averaging $W(h)$ over many initial realizations, we get the
probability of height at the first site being less than or equal to $h$,
i.e., $$ Prob_L(h_1 \le h) = \langle W(h) \rangle $$

The estimate of probability of the minimum configuration is obtained by
calculating the weight function \begin{equation}
 W_{min} = \prod_{\{{\cal C}_t \ne |11\dots11\rangle\}} F(t)
\end{equation} where the product is over all update steps $t$ required to
reach the minimum configuration. For different realizations, we get
different values of $W_{min}$ and, similarly as above, averaging over
different values by taking many realizations gives us the probability of
the minimum slope.
   
We illustrate this procedure by a simple example. Consider a rice pile
with $L=6$. At $t=0$, we have ${\bf u}(0) = \{1,2,3,4,5,6\}$, as all sites
are unstable. Also, at this stage $x_{max} =1$ for all sites. In this
case, the probability distribution of $y(1)$ is given by \begin{equation}
{\rm Prob}(y(1) \leq y| y(1) > p) = y^6/( 1-p^6), {\rm ~for~} p \leq y
\leq 1. \end{equation} This can be generated as follows: select a random
number $z$ uniformly distributed between $0$ and $1$. If $ z^{1/6}>p $,
put $y(1)= z^{1/6}$. If not, discard this value, and choose again. In this
case, we get $F(1)= 1 - p^6$. Then, we choose $j(1)$ as one of the sites
from ${\bf u}(0)$ at random, with equal probability.  Say, we get
$j(1)=2$. Then, we topple at this site and assign $x_{max}=z^{1/6}$ to
other sites with unstable $2$'s. Toppling makes sites $1$ and $3$
unstable, and we topple there as well. Whenever we topple at site with
slope $3$, we reset the $x_{max}$ at that site to $1$. Finally, toppling
at all unstable sites with $z =3$, we get the configuration with ${\bf
u}(1) = \{2,3,4,5,6\}$, and $x_{max}$ at all these sites is reset to $1$.  
So, now \begin{equation} {\rm Prob}(y(2) \leq y| y(1) > p) = y^5/( 1-p^5),
{\rm ~for~} p \leq y \leq 1, \end{equation} and $F(2)= 1 - p^5$. Now we
choose $j(2)$ at random from ${\bf u}(1)$, say $ j(2)=4$. Toppling at this
site induces toppling at other sites, and finally we get the configuration
of unstable sites ${\bf u}(2) = \{1,3,4,5,6\}$, and we have $x_{max}=1$ at
all these sites. We now generate the variable $y(3)$, which turns out to
have the same distribution as $y(2)$. Then we have $F(2) = 1 -p^5$. Now we
choose a site from ${\bf u}(2)$. and so on.

\subsection{Results.}

\begin{figure} \begin{center} \leavevmode
\includegraphics[width=9cm,angle=0]{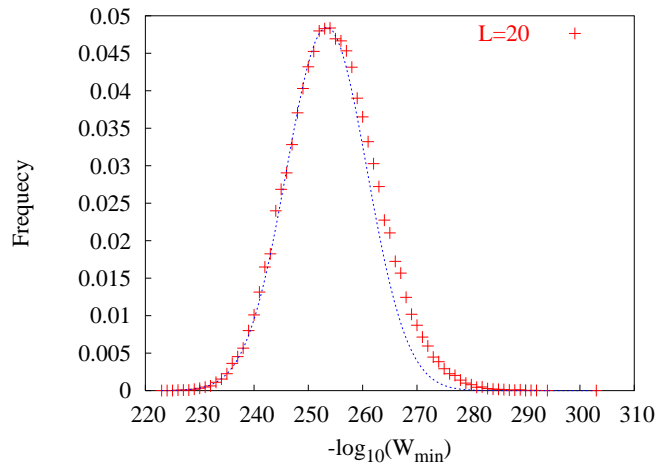} \caption{The frequency
distribution of $-\log_{10}(W_{min})$ taking bin size $=1$ and starting
with $10^6$ initial realizations for $L=20$. We fit the left tail of the
data points with a Gaussian distribution with mean $\mu \approx 253.4$ and
variance $\sigma^2 \approx 55$.} \label{error_bar} \end{center}
\end{figure}

We repeat the above procedure for many realizations and take the average
of logarithm of the weight $W(h)$. In Fig.\ref{error_bar} we 
have plotted the numerically obtained distribution of $\log W_{min}$ using 
$10^6$ initial realizations for $L=20$ . It has a peak at $\log W_{min}
\approx -253.4$ and decays rapidly about the peak value. We fit the data
point at the left of the peak value to a Gaussian distribution with mean
$\mu \approx -253.4$ and variance $\sigma^2 \approx 55$. It should be
noted here that our simulation cannot accurately estimate the
probabilities of near slope configurations for large $L$ and the
fractional error may be large, but the logarithm of the probabilities can
be estimated with reasonably small fractional error \cite{footnote2}.

\begin{figure} \begin{center} \leavevmode
\includegraphics[width=9cm,angle=0]{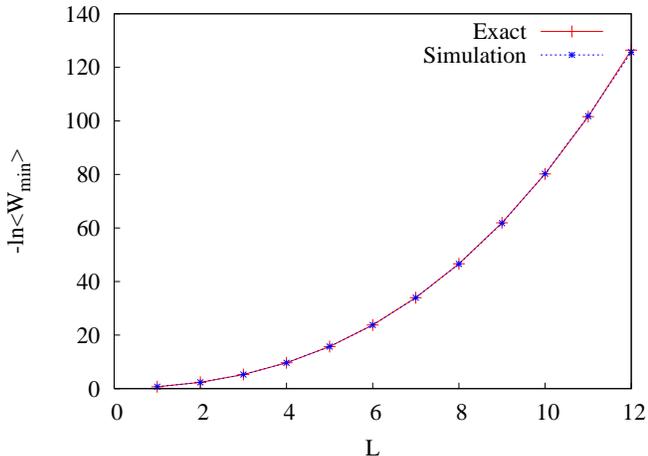} \caption{Exact
numerical calculation and the Monte Carlo simulation for $L \le 12$:
Logarithm (base $e$) of the probability of the minimum slope configuration
is plotted against the system size $L$ in the $1D$ Oslo ricepile model.
The data is averaged over $10^6$ realizations.} \label{exact_MC2}
\end{center} \end{figure}

As a check of our simulation algorithm, we calculated the probability of
the minimum slope configurations for small $L$ and the numerical values
match well with the values obtained from exact numerical calculation using
the method in section.3. For example, the probability of the minimum
slope, for $L=5$, is calculated to be $1.475 \times10^{-7}$ after
averaging the data over $10^6$ realizations and the value is correct upto
three significant digits. We have compared our results obtained from two
procedures, {\it i.e.}, the Monte Carlo simulation and exact numerical
calculation and plotted negative of logarithm of the probabilities against
$L$ in Fig.\ref{exact_MC2} for $L \le 12$.

For $h$ near $L$,  $W(h)$ is a product of approximately $L^3$ different
factors $F(t)$'s,  the logarithm of $W(h)$ is a sum of $L^3$ 
random variables.
While these variables $F(t)$'s are neither strictly independent nor they
are identically distributed, our simulation suggest that correlations
between these factors at different times are weak, so that we can expect
central- limit-theorem  like result to hold. Then $\log[W(h \simeq L)]$ 
may be expected
to be normally distributed with a mean and variance both proportional to
$L^3$ and $W(h)$ would be log-normally distributed. In fact, the 
experimentally obtained probability distribution function of $\log W$ 
shows significant deviations from gaussian  [ Fig.\ref{error_bar}]. 

Assuming that the distribution of the random variable  $X=-\ln [W(h \simeq
L)]$ has the form $\frac{1}{\sqrt{2\pi \sigma^2}}
\exp[-\frac{(X-\mu)^2}{2\sigma^2}]$, we get the probability of the minimum
slope equal to 
\begin{equation} 
Prob_L(h_1)  = \langle  W(h_1) \rangle = \langle e^{-X} \rangle  
\label{exp} 
\end{equation} 
Thus, if the gaussian approximation holds for the distribution,  
\begin{equation} \ln \langle W(h) \rangle \approx \langle \ln W(h) \rangle +
\frac{\sigma^2}{2} \label{exp1} \end{equation}
This estimate need not be
precise as the terms contribute significantly to $\langle W(h \simeq L)
\rangle$ are in the tail of the distributions and central limit theorem
need  not hold there.

\begin{figure} \begin{center} \leavevmode
\includegraphics[width=9cm,angle=0]{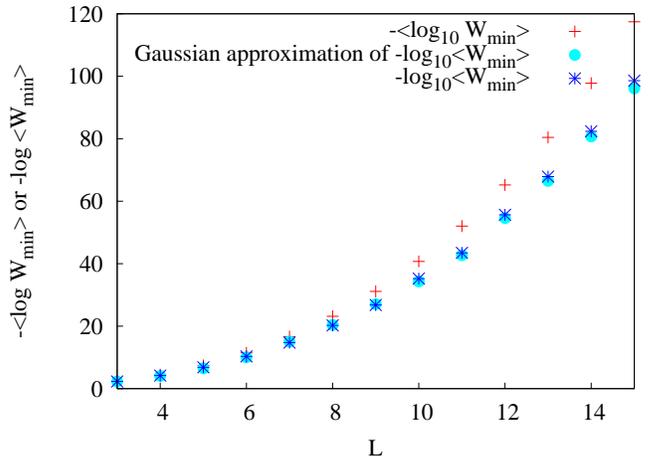} 
\caption{Comparision of  $\langle W_{min} \rangle$, as computed 
directly from the simulations (blue $\star$), and as  estimated by using the 
Gaussian approximation in Eq.\ref{exp1} (light blue bullet) for different 
system sizes $L$. Also shown is  $-\langle \log_{10}W_{min} \rangle$ 
(red $+$).}
\label{MC2_1112}
\end{center} \end{figure}

However our numerical estimate indicates that this approximation is indeed
very good. This is because the deviations from the gaussian behavior are 
stronger for smaller values of $W$, but these do not contribute much to 
$\langle W \rangle$.  For example, from the simulation for $L=15$, we get 
$\mu \approx -270$, $\sigma^2 \approx 95$, $\ln \langle W_{min} \rangle
\approx -229$. The Gaussian approximation to distribution of $\ln
W_{min}$ would give $\ln \langle W_{min} \rangle \approx - 223$.
In Fig.\ref{MC2_1112}, we have compared the actual values of $ 
log_{10} \langle W_{min} \rangle$  from the simulation with the estimate 
from the Gaussian
approximation for different values of the system size $L$. 

\begin{figure} \begin{center} \leavevmode
\includegraphics[width=9cm,angle=0]{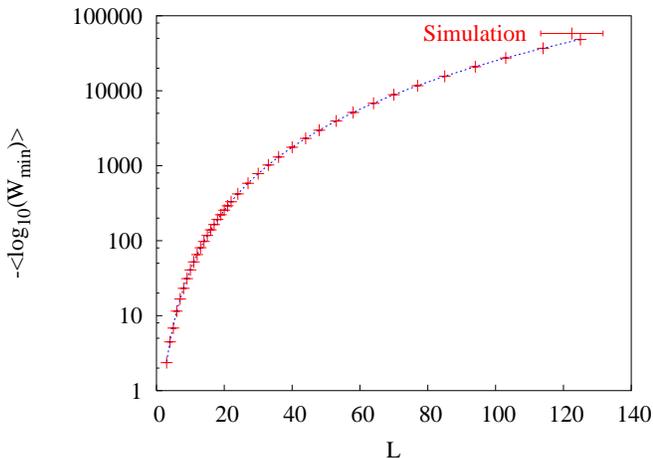} \caption{\small Monte
Carlo estimate of $\langle  \log W_{min} \rangle$   plotted against system 
size $L$ in the $1D$ Oslo ricepile model. The data points are fitted with
$ax^3+bx^2+cx$ where $a=0.0234$, $b=0.16$ and $c=0.1$.} \label{MC2}
\end{center} \end{figure}

In Fig.\ref{MC2} we have plotted negative of $\langle \log W_{min}
\rangle$ as a function of $L$ and fitted it with a curve $ax^3+bx^2+cx$.
We get a god fit for $a=0.0234\pm0.0001$, $b=0.16\pm0.05$ and
$c=0.11\pm0.2$.

\begin{figure} \begin{center} \leavevmode
\includegraphics[width=9cm,angle=0]{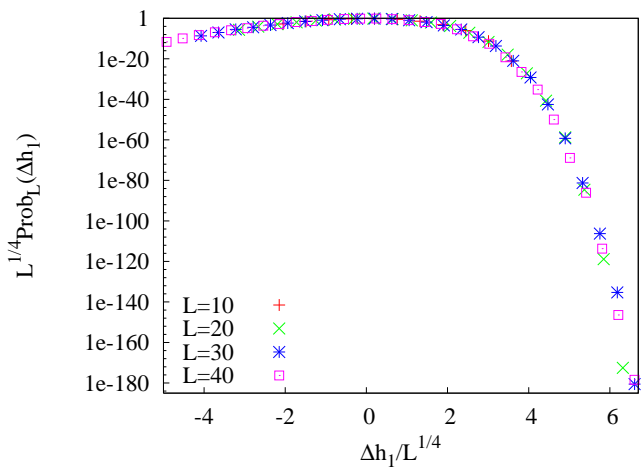} \caption{Scaling
collapse for the probability distribution of height at site $1$ for 
systems of size $L= 10, 20, 30, 40$.  $L^{1/4} Prob_L(\Delta h_1)$ has 
been plotted against   $(h_1-\bar{h}_1)/L^{1/4}$.
Each point is averaged over $10^6$ realizations.}
\label{h1_scaling_MC} \end{center} \end{figure}

\begin{figure} \begin{center} \leavevmode
\includegraphics[width=9cm,angle=0]{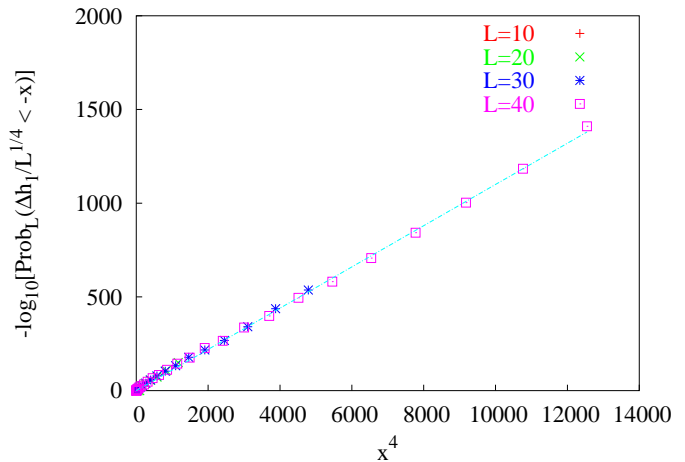}
\caption{$log_{10}[Prob_L(\Delta h_1/L^{1/4}< -x)]$, for $x>0$, has been 
plotted against 
the scaling variable $x^4$ where $\Delta h_1=(h_1-\bar{h}_1)$. The data
points are fitted with a straight line $0.11x$ and it shows that the
scaling function $g(-x)$ varies as $\exp(-\kappa x^4)$ for $x \gg 1$.}
\label{h1_scaling1_MC} \end{center} \end{figure}

Now we calculate the full probability distribution $Prob_L(h_1)$ of height
$h_1$ at site $1$. We take the average of this product over many
realizations and estimate $\log[Prob_L(h_1 \le h)]$ using the Gaussian
approximation. We thus determined  $Prob_L(h_1 =  h)$ for different $h$, 
for $L=10, 20, 30, 40$. The data is averaged over $10^6$ initial realizations.
In Fig.\ref{h1_scaling_MC}, we get a good scaling collapse by plotting
$L^{\omega} Prob_L(h_1 = h)$ against the scaling variable
$(h_1-\bar{h}_1)/L^{\omega}$ where $\omega \approx 1/4$. Therefore
$Prob_L(h_1 = h)$ a has scaling form $Prob_L(h_1) = L^{-1/4}
g[(h_1-\langle h_1\rangle)/L^{1/4}]$ at the central region as well as at
the tail. We see from the scaling plot the scaling function is highly
asymmetric about the origin.

Since the probability of minimum slope configurations varies as
$\exp(-\kappa L^3)$ for large $L$ and the height fluctuation $\Delta h_1=
(h_1-\bar{h}_1)$ scales with system sizes as $L^{1/4}$, the scaling
function $g(x)$ must vary as $\exp(-\kappa x^4)$. In Fig.\ref{h1_scaling1_MC}, 
we plot logarithm of $Prob_L(\Delta h_1/L^{1/4} < -x)$ versus $x^4$ for $x>0$
(i.e., fluctuations below average height). The data gives a reasonably good 
fit to a straight line with slope $0.11$.

\section{Summary}

We have done a Monte Carlo simulation using importance sampling to study
large deviations in the one dimensional Oslo ricepile model. We estimated
probabilities of large height fluctuations of the pile and these
probabilities are order $10^{-100}$ or even much less than this. We have
shown that logarithm of the probability of large negative height fluctuation
$\Delta h = -\alpha L$ varies as $- {\alpha}^{4} L^3$ for large $L$. We
also calculated numerically the full probability distribution of height of
the pile, and find that it has scaling form $L^{-1/4} g(\Delta
h/L^{1/4})$, with $\log [g(x)]$ varying as $ - x^4$ for large negative $x$.

\section{Appendix}

\begin{figure} \begin{center} \leavevmode
\includegraphics[width=9cm,angle=0]{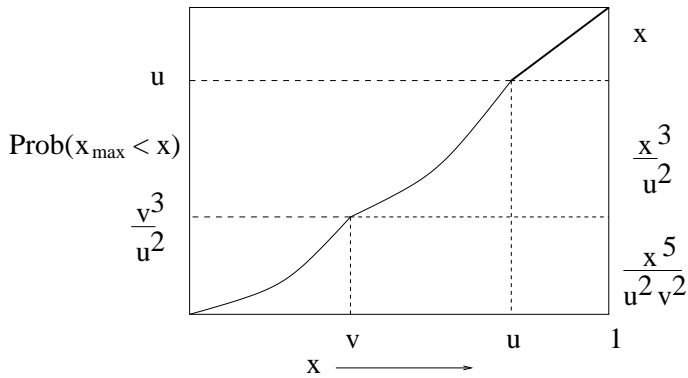} \caption{The probability
$Prob(x_{max} < x)$ of $x_{max}$ being less than $x$ versus $x$.}
\label{x_max} \end{center} \end{figure}

We illustate the procedure for generating the largest among $x(j,t)$'s, 
where $j \in {\bf u}$, using the
algorithm given here. Consider independent random variables $x_1$, $x_2$,
$x_3$, $x_4$, $x_5$ which are known to be uniformly distributed between
the respective intervals as given below,\\ \\ $
\mbox{~~~~~~~~~~~~~~~~~~~~~} 1. \mbox{~~} x_1 \in [0,1] \\
\mbox{~~~~~~~~~~~~~~~~~~~~~} 2. \mbox{~~} x_2 \in [0,u] \\
\mbox{~~~~~~~~~~~~~~~~~~~~~} 3. \mbox{~~} x_3 \in [0,u] \\
\mbox{~~~~~~~~~~~~~~~~~~~~~} 4. \mbox{~~} x_4 \in [0,v] \\
\mbox{~~~~~~~~~~~~~~~~~~~~~} 5. \mbox{~~} x_5 \in [0,v] $ \\ \\ where we
take $1 \ge u \ge v$ without loss of generality. Then the cumulative
probability distribution of $y$, the largest among these random numbers,
is given by \begin{eqnarray} \nonumber Prob(y \leq x) &=& x, {\rm ~for~ }
u \leq x \leq 1 \\ \nonumber
                &=& x^3/u^2,{\rm ~for~} v \leq x \leq u;  \\
                &=& x^5/(u^2 v^2),{\rm ~for ~} 0 \leq x \leq v.
\end{eqnarray} 
The distribution is drawn schematically in Fig. \ref{x_max}.
To generate a variable $y$ with this distribution, we use
the following algorithm: generate a number $z$ randomly between $0$ and
$1$. Then following cases are possible. \\ \\ 1. If $u < z \le 1$ we
choose the largest random number to be $y=z$ and the maximum is surely
$x_5$.  \\ 2. If $v^3/u^2 < z \le u$, we choose the largest random number
to be $y=(z u^2)^{1/3}$ and the maximum is chosen from $x_3$, $x_4$ and
$x_5$ with probability $1/3$ each. \\ 3. If $0 \le z \le v^3/u^2$, we
choose the largest random number to be $y=(z u^2 v^2)^{1/5}$ and the
maximum is chosen from $x_1$, $x_2$, $x_3$, $x_4$ and $x_5$ with
probability $1/5$ each. \\ \\ If the resulting value of $y$ is less than
$p$, the value is rejected, and the whole procedure is implemented afresh.  
Probability distributions for the maximum of more variables can be
obtained similarly.

\begin{thebibliography}{99}
\bibitem{footnote1} This time can be reduced to $\exp(L^2)$ if we store
the weights of $\exp(L)$ configurations to collect terms together.
\bibitem{footnote2} This is similar to the Monte Carlo in equilibrium
statistical physics, where thermodynamic quantities like free enrgy can be
estimated well, but not the partition function.
\end{thebibliography}
\end{document}